\documentclass[10pt,a4paper]{article}
\usepackage[latin1]{inputenc}
\usepackage{amsmath}
\usepackage{amsfonts}
\usepackage{amssymb}
\usepackage{makeidx}
\usepackage{graphicx}
\author{Sebastian Kokott}
\usepackage{bm}
\usepackage{amssymb}
\newcommand*{\vb}[1]{\bm{#1}}
\newcommand{\tx}[1]{\text{#1}}
\title{First-Principles Supercell Calculations of Small Polarons with Proper Account for Long-Range Polarization Effects}
\author{Sebastian Kokott$^1$, Sergey V Levchenko$^1$, \\ 
Patrick Rinke$^2$, and Matthias Scheffler$^{1,3}$}
\begin{document}
%\title[Small Polarons in Supercells]{First-Principles Supercell Calculations of Small Polarons with Proper Account for Long-Range Polarization Effects}

%\author{Sebastian Kokott$^1$, Sergey V Levchenko$^1$, 
%Patrick Rinke$^2$, and Matthias Scheffler$^{1,3}$}
\maketitle
\noindent\textit{\footnotesize
$^1$Fritz Haber Institute of the Max Planck Society, Faradayweg 4-6, 14195 Berlin, Germany \\
$^2$COMP/Department of Applied Physics, Aalto University School of Science, P.O. Box 11100, FI-00076 Aalto, Finland \\
$^3$University of California at Santa Barbara, CA 93106, USA }
%\ead{kokott@fhi-berlin.mpg.de}

\section*{Abstract}
We present a density functional theory (DFT) based supercell approach for modeling small polarons with proper account for the long-range elastic response of the material. Our analysis of the supercell dependence of the polaron properties (e.g., atomic structure, binding energy, and the polaron level) reveals long-range electrostatic effects and the electron-phonon interaction as the two main contributors. We develop a correction scheme for DFT polaron calculations that significantly reduces the dependence of polaron properties on the DFT exchange-correlation functional and the size of the supercell in the limit of strong electron-phonon coupling. Using our correction approach, we present accurate all-electron full-potential DFT results for small polarons in rocksalt~MgO and rutile~TiO$_2$.

%\noindent{\it Keywords\/}: density functional theory, polarons, finite-size correction, electron-phonon coupling \\

%\submitto{\NJP}

%\maketitle

\section{Introduction}
The electron-phonon (el-ph) interaction is fundamental to materials. It mediates, for example, the excitation of phonons in response to electronic excitations, which is especially pronounced in polar materials. These phonon excitations can stabilize a lattice distortion around a single excess charge (electron or hole). The excess charge and its accompanying lattice distortion are then referred to as a quasiparticle or more specifically as polaron. The formation and migration of polarons determine the properties of functional materials, such as their catalytic\cite{Henderson2011,Dohnalek2010} and photovoltaic\cite{Zhu2015} behavior. The direct observation of polarons in experiments, e.g. with electron-paramagnetic resonance \cite{Yang2013}, UV/IR spectroscopy \cite{Sezen2015}, or scanning tunneling microscopy or spectroscopy \cite{Setvin2014} is difficult, and computational studies are required to interpret the experimental data correctly. In this work, we develop a new method that addresses challenges faced in computational modeling of small polarons in materials with strong electron-phonon coupling, in particular in oxides, with density-functional theory (DFT).

\begin{figure}[b]
\includegraphics[width=0.9\textwidth]{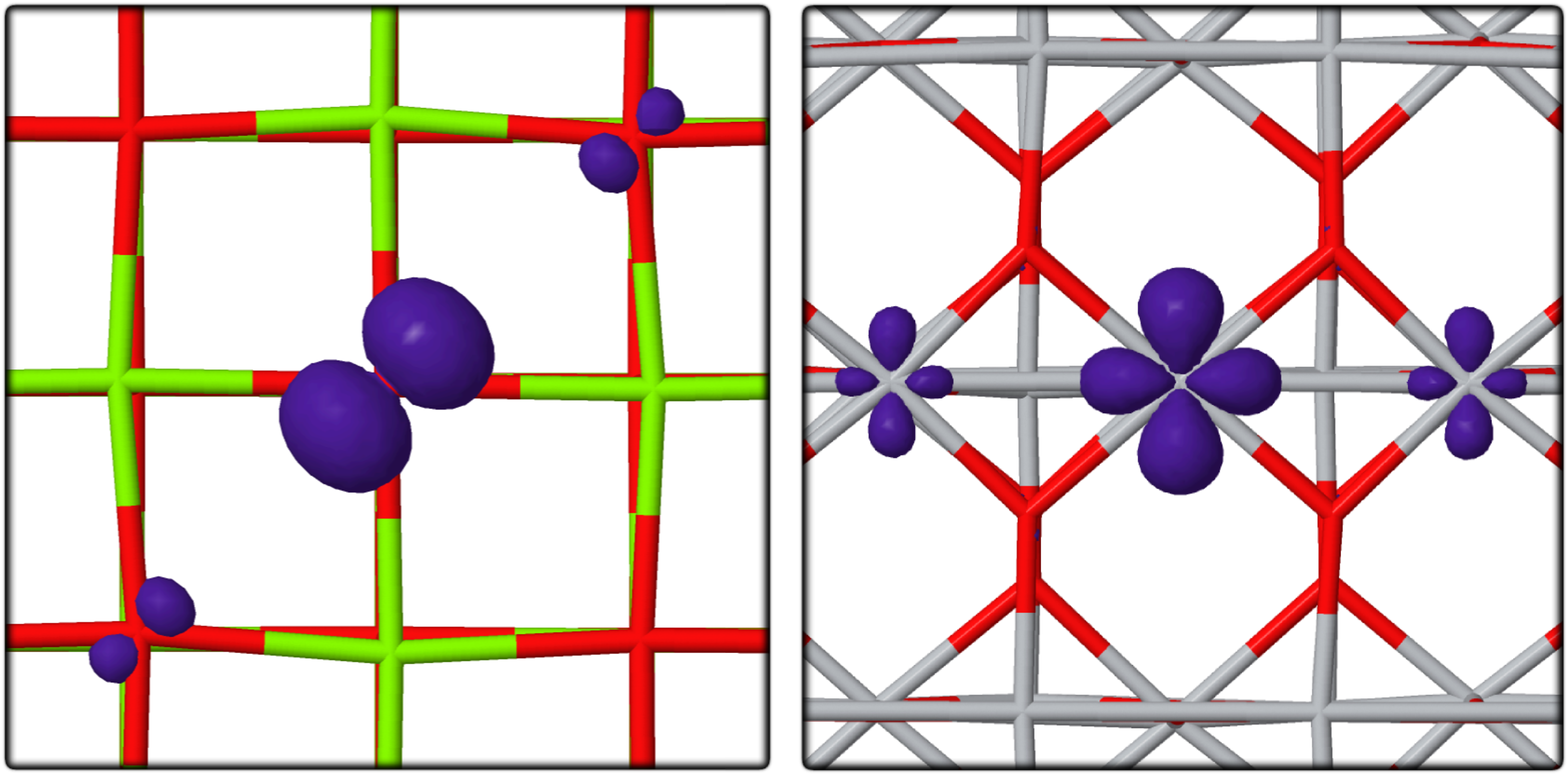}
\caption{The eigenstate densities for the hole polaron in MgO (left panel) and the electron polaron in TiO$_2$ (right panel) illustrate the strong localization of the excess charge. The isosurfaces encompass 0.95 of the polaron density. Red nodes  represent oxygen, green magnesium, and silver titanium atoms.}
\end{figure}

Polarons can be classified by their size as quantified by the extent of their total wave function (electrons and ions). Large polarons are delocalized over several unit cells and usually appear, if the el-ph interaction is weak. Such polarons were first investigated by Fr\"ohlich\cite{Froehlich1954}. In contrast, small polarons are mainly localized on one atomic site and form when the el-ph interaction is strong. Intermediate polarons \cite{Sezen2015} cover the size range in between. Pioneering work on small polarons was performed by Holstein\cite{Holstein1959}. He took only short-range interactions into account and identified the Fr\"ohlich coupling constant $\alpha_\tx{Fr\"ohlich}$ \cite{Froehlich1954} as good indicator for the el-ph interaction strength. Oxides fall into the intermediate to strong coupling regime, i.e., $\alpha_\tx{Fr\"ohlich}>1$. For instance, for MgO $\alpha_\tx{Fr\"ohlich}$ is 4.4 and for rutile TiO$_2$ 2.2. We therefore expect small polaron formation in both of these oxides. However, since MgO and TiO$_2$ are strongly ionic, the distortion of the ionic lattice can be long-ranged in violation of the polaron classification scheme. Such polarons in which the excess charge is localized, but the lattice distortion delocalized are referred to as small Fr\"ohlich polarons.

To describe small Fr\"ohlich polarons accurately in computational materials modeling, both long- and short-range interactions have to be treated appropriately. How to accomplish this task in DFT calculations that employ supercells, whose extend is typically smaller than the ionic lattice distortions, is the subject of this paper. Since small polarons can be regarded as a special type of a point defect, our study is also useful for point defect calculations of this type, which have so far eluded a reliable theoretical treatment. 

The paper is structure as follows. In Section~\ref{elastic_long_range} we derive the electrostatic and the electron-phonon contributions to the elastic long-range response of a material to a localized excess charge. We will then use this derivation to develop a correction scheme that removes artificial interactions from the supercell approach to obtain polaron properties in the dilute limit. In light of our new understanding, we analyze shortcomings of previous polaron approaches in Section~\ref{polaron_approaches}. In Section~\ref{sectio:numerical_results}, we demonstrate the efficiency of our approach  for hole polarons in MgO and electron polarons in rutile TiO$_2$.

\section{Elastic long-range behavior}
\label{elastic_long_range}

DFT in combination with the supercell approach has become the method of choice for the \textit{ab inito} calculation of point defects in solids. However, the supercell approach suffers from finite-size effects, especially for charged defects. These finite-size effects include the interaction of the excess charge with its periodic images, with the compensating constant background charge introduced to keep the unit cell neutral, and with the periodic constraint on the atomic relaxation. To overcome these finite-size limitations, two strategies are commonly used: (a) extending the supercell and extrapolating to the dilute limit based on a scaling law, or (b) applying an \textit{a posteriori} correction. For (a) only general knowledge about the size dependence is necessary. For example, the formation energy of a charged defect in the bulk as a function of the supercell size $L$ ($L=\Omega^{1/3}$, where $\Omega$ is the supercell volume) can be written as an inverse powerlaw:
\begin{equation}
E(L) = E(\infty) + a_1 \frac{1}{L}+ a_3 \frac{1}{L^3},
\label{scaling_law}
\end{equation}
where $E(\infty)$ is the formation energy in the dilute limit. This scaling law was derived by Makov and Payne\cite{Makov1995}. The disadvantage of this procedure is that at least three supercell calculations of increasing size are needed to fit $E(\infty)$ in Eq.~\eqref{scaling_law}, which is computationally very demanding, especially if atomic relaxations are included\footnote{Alternatively, it is possible to embed the central region in the pristine crystal via a Green's function approach (see e.g. Ref.~\cite{Scheffler1985})}.

Conversely, approach (b) requires an appropriate physical model for the long-range interactions in the solid. If only the electronic response to the excess charge is considered, its long-range contribution to the energy is described by a term proportional to $1/ \epsilon_\infty r$ (e.g. Ref.~\cite{Freysoldt2009}). However, if the ionic response cannot be neglected, the problem becomes challenging, and so far this case has not been solved. It has been suggested that the long-range elastic contribution is similar to the electronic one, but with the high-frequency dielectric constant $\epsilon_\infty$ replaced by the static one $\epsilon_0$, i.e., the long-range potential behaves classically like $1/\epsilon_0 r$. However, corrections based on this assumption generally overestimate $E(\infty)$, especially for vacancies. This overestimation has two reasons. First, the aforementioned long-range behavior is a crude approximation, neglecting all details of the underlying phonon structure. Second, the short-range screening of the excess charge depends on the el-ph coupling, which determines the interplay of the electronic and ionic responses. Screening can then be more efficient than in the static limit. In the following we analyze the screening effects in detail and show that only in the strong-coupling limit of the el-ph interaction the substitution of $\epsilon_\infty$ with $\epsilon_0$ is a good approximation.

\begin{figure}

\includegraphics[width=\textwidth]{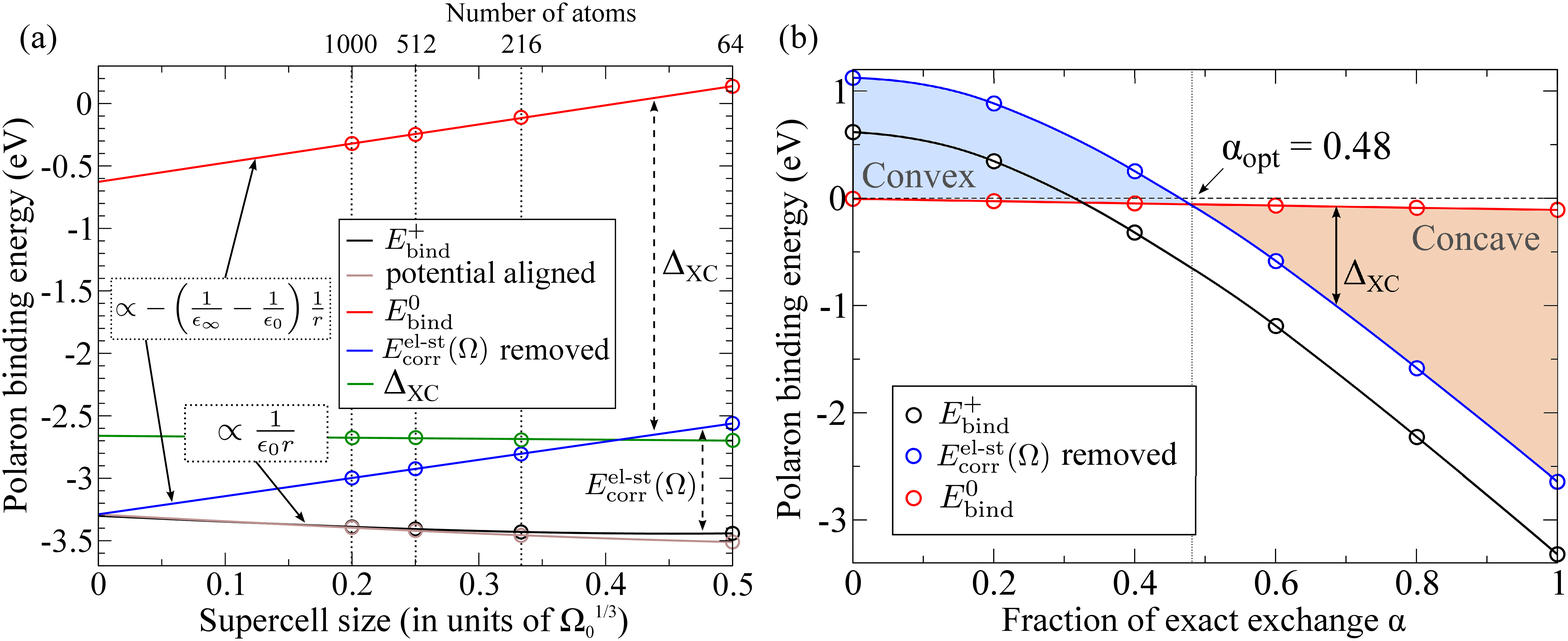}
\caption{(a) Supercell dependence of the polaron binding energy  for HSE($\alpha$=1). The atomic positions are fully relaxed for each supercell size and for each functional. (b) Dependence of the polaron binding energies on the fraction of exact exchange. The fixed geometry of the $3\times3\times3$ supercell from panel (b) is used and binding energies for different fraction of exact exchange are calculated.}
\label{fig:polaron_alpha_1}
\end{figure}

We start by splitting the long-range elastic potential $V^\tx{lr}$ into the electron-phonon interaction $ V^\tx{lr}_\tx{el-ph}$  and electrostatic potential $V^\tx{lr}_\tx{el-st}$:
\begin{equation}
V^\tx{lr} = V^\tx{lr}_\tx{el-st} + V^\tx{lr}_\tx{el-ph} \,  .
\label{lr_potential}
\end{equation}
$V^\tx{lr}_\tx{el-st}$ is generated by the charge density $\rho_d (\vb{r})$ of the localized excess charge. The Fourier transform of $V^\tx{lr}_\tx{el-st}$ is then given by:
\begin{equation}
V^\tx{lr}_\tx{el-st}(\vb{k}) = 2\pi \frac{\rho_d (\vb{k})}{\vb{k}^T\vb{\epsilon}_\infty \vb{k}} \, ,
\label{elst_potential}
\end{equation} 
where $\rho_d (\vb{k})$ is the Fourier transform of $\rho_d (\vb{r})$. 
%and we have taken a factor 1/2 into account for double counting. 

To obtain a corresponding expression for $V^\tx{el-ph}$ we first have to introduce additional assumptions. First, we will focus on ionic crystals. Second, we only consider the interaction of an electron with a single phonon at a time, neglecting higher-order contributions. Third, we assume that the adiabatic approximation (factorization of the electron and phonon wave functions) and strong electron-phonon coupling limit ($\alpha_\tx{Fr\"ohlich}\gtrsim 4$) are applicable. With these assumptions the long-range part of $V^\tx{el-ph}$ reduces to \cite{Devreese2016}:
\begin{equation}
V^\tx{lr}_\tx{el-ph} (\vb{k}) =  -\sum_{\nu} \frac{1}{ \hbar \omega_{\vb{k}\nu}} \left\vert g^\tx{lr}_\tx{el-ph} (\vb{k}\nu)\right\vert^2 \rho_d (\vb{k}) \, .
\label{elph_potential}
\end{equation}
The potential in Eq.~\eqref{elph_potential} is attractive, lending further stabilization to the polaron. An analytic expression for $ g^\tx{lr}_\tx{el-ph}$ was recently derived by Verdi and Guistino\cite{Verdi2015}:
\begin{equation}
g^\tx{lr}_\tx{el-ph} (\vb{k}\nu) = i 4\pi e \sum_{\kappa} \left( \frac{\hbar}{2NM_\kappa \omega_{\vb{k}\nu}} \right)^{1/2} \frac{\vb{k}^T \vb{Z}^\ast_\kappa \vb{e}_{\kappa\nu}(\vb{k})}{\vb{k}^T\vb{\epsilon}_\infty \vb{k} }
\label{elph_vertex}
\end{equation}
where $\nu$ labels the phonon mode,  $\omega_{\vb{k}\nu}$ is the corresponding phonon frequency of ion $\kappa$ with mass $M_\kappa$.  $\vb{Z}^\ast_\kappa$ is the Born effective charge tensor  and $\vb{e}_{\kappa\nu}(\vb{k})$ are the phonon eigenvectors of the dynamical matrix. 

Equations~\eqref{elph_potential} and~\eqref{elph_vertex} describe the scattering of all phonon modes with $\rho_d$. Thus, the long-range behavior of the el-ph interaction depends on the phonon structure across the entire phonon Brillouin zone, and the elastic behavior is not captured by the classical $1/\epsilon_0 r$ limit. If we only consider the interaction of $\rho_d$ with a single dispersion-less longitudinal-optical mode $\omega_{LO}$, we recover the limit of the Fr\"ohlich electron-phonon interaction in the strong-coupling limit, which was first investigated by Pekar\cite{Pekar1946}. With the Fr\"ohlich matrix element (for the anisotropic case we refer to \cite{Fock1975}):
\begin{equation}
g^F (k) = i e  \left[2 \pi \hbar \omega_{LO}\left( \frac{1}{\vb{k}^T\vb{\epsilon}_\infty \vb{k}}-\frac{1}{\vb{k}^T\vb{\epsilon}_0 \vb{k}}\right)\right]^{1/2}
\end{equation}   
we obtain the potential:
\begin{equation}
V^\tx{lr}_\tx{el-ph} (\vb{k}) = - 2\pi  \frac{\rho_d (\vb{k})}{\vb{k}^T\vb{\epsilon}_\infty \vb{k}} + 2\pi \frac{\rho_d (\vb{k})}{\vb{k}^T\vb{\epsilon}_0 \vb{k}} \, .
\label{pekar_potential}
\end{equation}
Upon substituting Eq.~\eqref{pekar_potential} and Eq.~\eqref{elst_potential} into Eq.~\eqref{lr_potential}, we finally arrive at the classical limit of a screened potential for a localized charged distribution in an anisotropic medium:
\begin{equation}
V^\tx{lr}(\vb{k}) = 2\pi  \frac{\rho_d (\vb{k})}{\vb{k}^T\vb{\epsilon}_0 \vb{k}} \, .
\label{final_elastic}
\end{equation}
The el-ph potential given by Eq.~\eqref{pekar_potential} is an upper bound and, consequently, Eq.~\eqref{final_elastic} is also an upper bound. This explains why any correction based on Eq.~\eqref{final_elastic} overestimates the actual limit. We find that, despite the approximations we made, $V^\tx{lr}(\vb{k})$ in Eq.~\eqref{final_elastic} is still appropriate for polarons in the intermediate coupling regime ($ 1 \lesssim \alpha_\tx{Fr\"ohlich} \lesssim 4$). Vice versa, our derivation illustrates ways to improve the long-range model for polarons and charged point defects, if needed, since all assumptions are clearly defined. 

Based on the knowledge of the long-range behavior, the errors due to finte size of the supercell can be corrected using {\em a posteriori} methods, such as  the method of Freysoldt \textit{et al}.\cite{Freysoldt2009}. For technical details we refer to \cite{Freysoldt2009,Freysoldt2011}. Generalizing the Freysoldt method to an arbitrary interaction potential $V(\vb{r})$ and anisotropic media (in the standard approach of Freysoldt {\em et al}., $V(\vb{r}) = 1/\epsilon_{\infty} r$), the correction for the interaction energy is obtained as the difference between the energy of the artificial lattice of charged defects, $E_\tx{latt}$, and the energy of an isolated defect, $E_\tx{iso}$: 
\begin{eqnarray}
 E_{corr}(\Omega) &&= E_\tx{latt}(\Omega) - E_\tx{iso} \nonumber \\
 &&=\frac{1}{\Omega}\sum_{{\bf G}\neq {\bf 0}} V(\vb{G}) q_m(\vb{G}) 
 - \frac{1}{(2\pi)^3} \int V(\vb{k}) q_d(\vb{k}) \tx{d}\vb{k}\,\, ,
 \label{eq:ecorr}
\end{eqnarray} 
where $V$ can be $V^\tx{lr}_\tx{el-ph}$, $V^\tx{lr}_\tx{el-st}$ or the sum of both, and $q_d(\vb{k})$ is the Fourier transform of the excess charge distribution, and $q$ is the total charge. Taking into account the alignment terms $q\Delta V$~\cite{Freysoldt2009,Freysoldt2011} A summary of the Freysoldt \textit{et al.} correction scheme including the meaning of the alignment terms can be found in the Appendix~\ref{app:Freysoldt}, the corrected energy is obtained as:
\begin{equation}
E(\infty) = E(\Omega) -E_{corr}(\Omega) + q\Delta V\, .
\label{eq:total_corr}
\end{equation}
Having derived the correction for the elastic contribution, we can apply it to the polaron problem and investigate the effects of the two parts in Eq.~\eqref{lr_potential} separately.

\section{The Polaron in a supercell}
\label{polaron_approaches}
\subsection{The charged supercell}
An important property of a polaron is its binding energy 
\begin{equation}
E_\text{bind}^\pm = E^\text{polaron} (N\mp1) - E^\text{perf} (N\mp1)
\label{BEPo}
\end{equation}
where the energies have not been corrected for finite-size effects, yet. The plus in $E_\text{bind}^\pm$ corresponds to electron removal (hole polaron), while the minus sign corresponds to electron addition (electron polaron). $E^\text{polaron}$ is the total energy of the distorted system (polaron geometry), $E^\text{perf}$ the total energy of the undistorted system. The number of electrons in the system are given in parenthesis, with $N$ corresponding always to the neutral system. A negative $E_\text{bind}^\pm$ indicates an energy gain and a stable (self-trapped) polaron. 

In the following we focus on the hole polaron for brevity, since only small adjustments of the formalism are needed for the electron polaron case. The simplest way to calculate the polaron binding energy is straightforward: in Eq.~\eqref{BEPo} $E^\text{polaron} (N\mp1)$ is computed with DFT and full structure relaxation in the charged supercell. To ease the system out of possible high symmetry configurations an initial symmetry-breaking distortion might have to be applied. Finite-size effects are expected to be small, since  the elastic long-range interaction falls off with $1/\epsilon_0 r$ and the static dielectric constant $\epsilon_0$ is usually large ($\gtrsim$10) for ionic crystals (however, as demonstrated and explained below, the dependence of the polaron binding energy defined by Eq.~\eqref{BEPo} on the approximations in the exchange-correlation functional is strong). The supercell dependence of $E_\text{bind}^+$ for MgO is shown in  Fig.~\ref{fig:polaron_alpha_1}, panel a, where we used HSE06 hybrid functional\cite{Heyd2003,Heyd2006} with the fraction of exact exchange $\alpha=1$ [denoted HSE06($\alpha=1$); see Section \ref{sectio:numerical_results} for more computational details]. We find a small hole polaron mainly localized at the central oxygen atom. The displacements of the nearest neighbors are of the order of 0.1~\AA~and decaying fast away from the center. The shape of the excess charge density distribution is $p$-like. For sufficiently large supercells, when the long-range regime is valid, the dependence of the binding energy on the supercell size $L$ becomes $1/\epsilon_0 L$. From the slope of $E_\text{bind}^+(1/L)$ at $1/L = 0$ we obtain $\epsilon_0=10.32$, in good agreement with the experimental static dielectric constant for MgO $\epsilon_0=9.8$. 

Next, we calculate the correction for the artificial electrostatic interaction due to the periodic arrangement of the holes and their interaction with the constant background, using Eq.~\eqref{eq:ecorr} with the potential$1/\epsilon_\infty r$. To model the excess charge density $\rho_d (\vb{r})$ needed here and for following finite-size corrections, we fit the envelope of the KS eigenstate density (decays exponentially for a localized state) with an exponential function $\rho_\tx{model}=A\, \tx{exp}(-|r-r_0|/\gamma)$, where $A$ is a normalization constant, $r_0$ is the center of the polaron, and $\gamma$ the fitting parameter corresponding to the polaron radius (cf. App.~\ref{app:pekar_polaron}). Additionally, we calculated the alignment term $\Delta V$ in Eq.~\eqref{eq:total_corr} between the charged, neutral, and model (i.e., including the model excess charge density compensated by a constant background charge) systems following the approach outlined in Ref.~\cite{Freysoldt2011}. After this correction, according to Eq.~\eqref{lr_potential} the remaining contribution is due to the long-range electron-phonon interaction. This contribution is shown by the blue line in Fig.~\ref{fig:polaron_alpha_1}, panel a. The line is almost perfectly straight, and the slope is equal to $\epsilon_\infty^{-1}-\epsilon_0^{-1}$, where $\epsilon_0=10.32$ is taken from the fit of $E_\tx{bind}^+$ presented above, and  $\epsilon_\infty=2.4$ is obtained from an independent calculation~\footnote{The dielectric constant $\epsilon_\infty$ was obtained by fitting Eq.~\eqref{scaling_law} for the unrelaxed doubly charged oxygen vacancy in MgO for three different supercell sizes.}. This analysis explains the role of different long-range interactions in Eq.~\eqref{lr_potential} in the supercell dependence of polaron properties.

Thus, the approximations in Eq.~\eqref{pekar_potential} work well for MgO, which is expected since it has only one longitudinal optical phonon mode, strong el-ph coupling, and is an isotropic material. However, we find that the polaron binding energy defined by Eq.~\eqref{BEPo} is extremely sensitive to the approximations in the exchange-correlation functional. Figure~\ref{fig:polaron_alpha_1}, panel b, shows the dependence of the binding energy on the fraction of exact exchange $\alpha$ in the HSE06 functional. Within a small range $\pm 0.05$ of $\alpha$ around the standard value (0.25) the binding energy changes by about 0.5 eV. This leads to a qualitative change in small polaron stability, from a stable self-trapped polaron (negative binding energy) to an unstable small polaron (positive binding energy). This strong functional dependence makes even a qualitative assessment of the existence of a self-trapped polarons impossible. Several approaches have been suggested in the literature for determining the correct or at least optimal value of $\alpha$\cite{Atalla2013,Atalla2016,Lany2011,Skone2014,Dauth/etal:2016,Refaely2011,Stein2010}. Here we focus on \textit{restoring the IP theorem}\cite{Lany2011} as a consistent \textit{DFT-based} solution of the problem.

In (exact) DFT within the scope of Kohn-Sham (KS) scheme the vertical ionization potential $IP$ should be equal to the negative of the highest occupied KS state energy $\varepsilon_\text{ho}$ in the system:
\begin{equation}
IP^\ast = E(N-1) - E(N) = -\varepsilon_\text{ho}(N)\, ,
\label{IP_theorem_mol}
\end{equation} 
where $E(N-1)$ and $E(N)$ are total energies of the ionized and neutral system, respectively.
In this work we refer to this relation as IP-theorem, but it is also know as HOMO-I condition\cite{Atalla2013} or Generalized Koopmans' Theorem\cite{Lany2011}, and is directly related to the straight-line dependence of the total energy on occupation of the highest-occupied state\cite{Perdew1982} or the fact that the position of $\varepsilon_\text{ho}$ is independent on its occupation. Equation~\eqref{IP_theorem_mol} is always correct for any extended (delocalized) state, as was already pointed out by Janak (1978) and extended to the case of the generalized KS scheme by Perdew {\em et al}.\cite{Perdew2017}. However, for a given density-functional approximation (DFA)  Eq.~\eqref{IP_theorem_mol} does not necessarily hold, if the orbital is localized, unless the satisfaction of the straight-line condition is explicitly included in the design of the functional. The deviation from the straight line $\Delta_\text{XC}(\alpha)$ is described by two contributions to Eq.~\ref{IP_theorem_mol}:
\begin{equation}
E(N-1) - E(N) = -\varepsilon_{\rm ho}(N) + \Delta_\text{XC}(\alpha), \, \Delta_\text{XC}=\Pi+\Sigma,
\label{IP+nK}
\end{equation}
with the self-interaction error $\Pi$ causing a convex curvature of the total energy as a function of occupation, and the orbital relaxation $\Sigma$ a concave curvature. The optimal $\alpha=\alpha_{\rm opt}$ minimizing the self-interaction error~\cite{Atalla2016,Lany2009} is then determined from the condition $\Delta(\alpha_{\rm opt})=0$.

The straight-line theorem (Eq.~\eqref{IP_theorem_mol}) was originally proven for finite systems, and transferring it to a solid with periodic boundary conditions needs special care. For any finite supercell with volume $\Omega$ the energy of the artificial electrostatic interactions due to the periodic arrangement [$E_\tx{corr}^\tx{el-st}(\Omega)$, obtained using Eq.~\eqref{eq:ecorr} with potential from Eq.~\eqref{elst_potential}] has to be removed from $E(N-1)$:
\begin{equation}
E(N-1) - E_\tx{corr}^\tx{el-st}(\Omega) - E(N) = -\varepsilon_\tx{ho}(N) + \Delta_\text{XC}(\alpha)\, ,
\label{IP_theorem_pbc}
\end{equation} 
since it would only vanish in the limit of an infinite supercell. Combining Eq.~\eqref{IP_theorem_pbc} and Eq.~\eqref{BEPo}, we get:
\begin{eqnarray}
E_\tx{bind}^+ = E_\tx{bind}^0 + E_\tx{corr}^\tx{el-st} + \Delta_\tx{XC}(\alpha)\, ,
\label{BEPo_neutral}
\end{eqnarray}
where $E_\tx{corr}^\tx{el-st}$ stands for the artificial electrostatic interaction energy for the distorted geometry, since for the perfect geometry it is zero.
The quantity 
\begin{eqnarray}
E_\tx{bind}^0 =  \Delta E^\tx{polaron} -E_0
\label{SIF_BEPo}
\end{eqnarray} 
is calculated using only {\em neutral} unit cells, with the energy of distortion from perfect to polaronic geometry $\Delta E^\tx{polaron}=E^\tx{polaron} (N) - E^\tx{perf} (N)$ and the polaron level energy with respect to the VBM $E_0=\varepsilon_\tx{ho}^\tx{polaron} (N) -\varepsilon_\tx{VBM}^\tx{polaron} (N)$. According to Eq.~\eqref{BEPo_neutral}, when $\Delta_\tx{XC}(\alpha)$ is zero, $E_\tx{bind}^0$ represents the polaron binding energy corrected for the artificial electrostatic interaction.

\begin{figure}
\includegraphics[width=\textwidth]{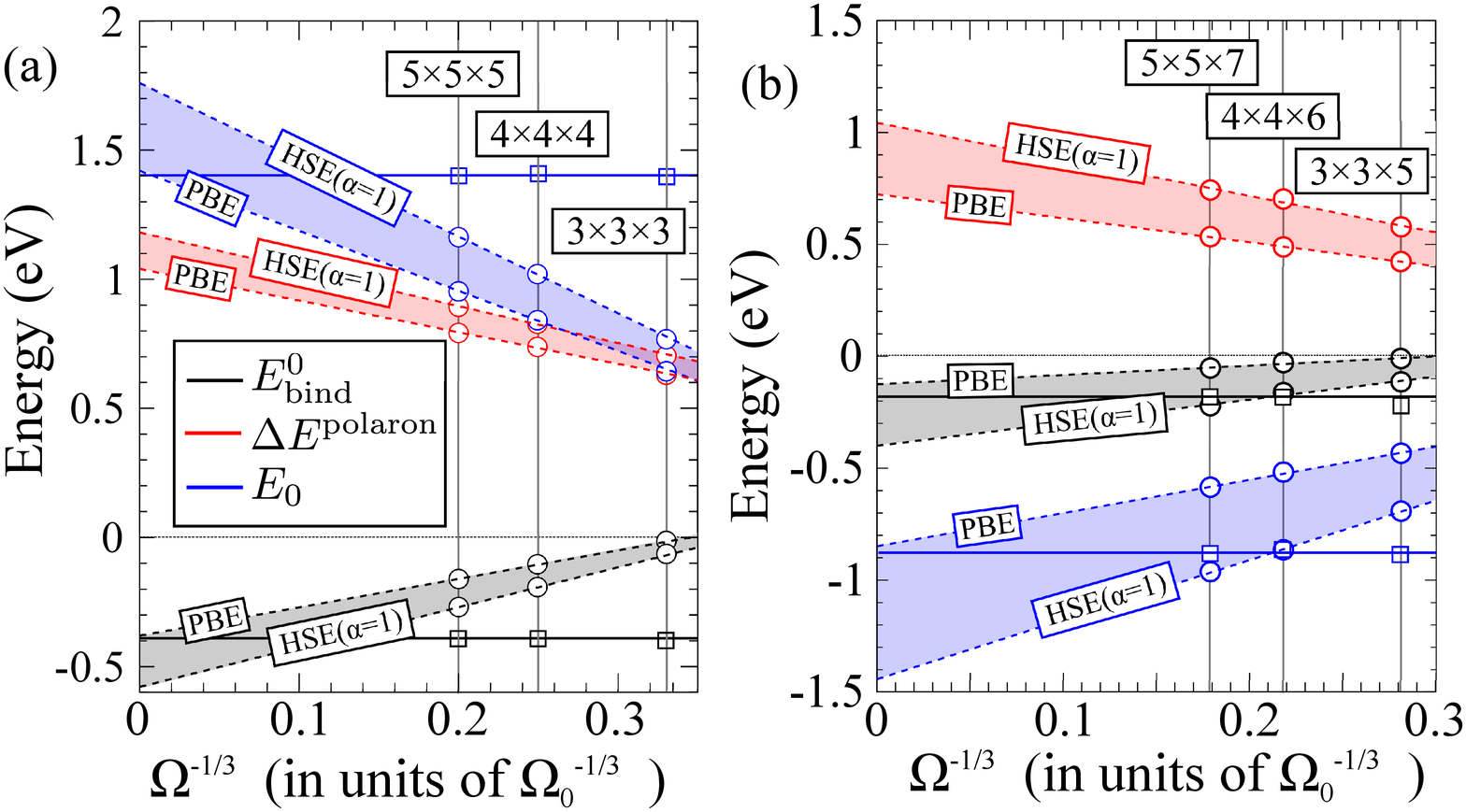}
\caption{The polaron binding energy $E^\tx{IP}_\tx{bind}$ (Eq.~\ref{SIF_BEPo}), the polaron KS level $E_0$ with respect to band edge, and the  relaxation energy $\Delta E^\tx{polaron}$ for MgO (left panel) and rutile TiO$_2$ (right panel). The x axis is given in units of the cubic root of the unit cell volume $V_0$. The PBE polaron binding energies corrected for the finite-size effects are shown by square symbols. The solid lines show linear least-squares fit for different energy components and DFT approximations. For all supercells the atoms are relaxed according to the approach described in the text.}
\label{figure_2}
\end{figure}

$E_\tx{bind}^0$ is shown in Fig.~\ref{fig:polaron_alpha_1}, panel a, as red line (top-most line), where the optimized polaron geometry of the charged supercell is used. Note that, despite including only quantities calculated using neutral unit cells, $E_\tx{bind}^0$ has a strong dependence on the unit cell size. As discussed below, this dependence is due to the interaction of the ionic relaxations in different unit cells. Taking the difference between the blue and the red lines, we find that the exchange-correlation error $\Delta_\tx{XC}$ is practically independent on the unit cell size (green line in Fig.~\ref{fig:polaron_alpha_1}, panel a), starting from the smallest supercell with 64 atoms we have considered. 
This implies that $ \Delta_\text{XC}(\alpha)$ could be calculated even in the smallest supercell and then removed in any larger supercell. In order to obtain the optimized $\alpha=\alpha_\tx{opt}$ we have to remove $E_\tx{el-st}(\Omega)$ from the binding energy $E^+_\tx{bind}$ and determine the intersection with $E_\tx{bind}^0$. The result is shown in Fig.~\ref{fig:polaron_alpha_1}, panel b, and we obtain $\alpha_\tx{opt}=0.48$. Since the dependence on $\alpha$ is not linear, at least three different values of $\alpha$ have to be calculated to estimate $\alpha_\tx{opt}$. Additionally for each value of $\alpha$ the dielectric constant $\epsilon_\infty$ has to be calculated. Thus, the simulation of the polaron in a charged supercell is computationally demanding, since it is extremely sensitive to the underlying functional. In the next subsection we demonstrate an approach to overcome this problem.

\subsection{The neutral supercell}

As mentioned above, $E_\tx{bind}^0$ in Eq.~\eqref{BEPo_neutral} is equal to the polaron binding energy corrected for the artificial electrostatic interaction, only when $\Delta_\tx{XC}(\alpha)$ vanishes. However, similar to previous work~\cite{Sadigh2015,Zawadzki2011} we find that $E_\tx{bind}^0$ is far less sensitive to the underlying functional than $E_\tx{bind}^+$, as can be seen for MgO in Fig.~\ref{fig:polaron_alpha_1}, panel b. The same is true for TiO$_2$, but the remainig dependence is larger than for MgO (see Fig.~\ref{figure_2}). This has an interesting implication: $E_\tx{bind}^0$ is the polaron binding energy with most of the exchange-correlation error removed (because for an exact functional $E_\tx{bind}^0$ is equal to the polaron binding energy). The reason for the insesitivity of $E_\tx{bind}^0$ on the functional remains unclear~\cite{Sadigh2015}.

\begin{figure}
\includegraphics[width=\textwidth]{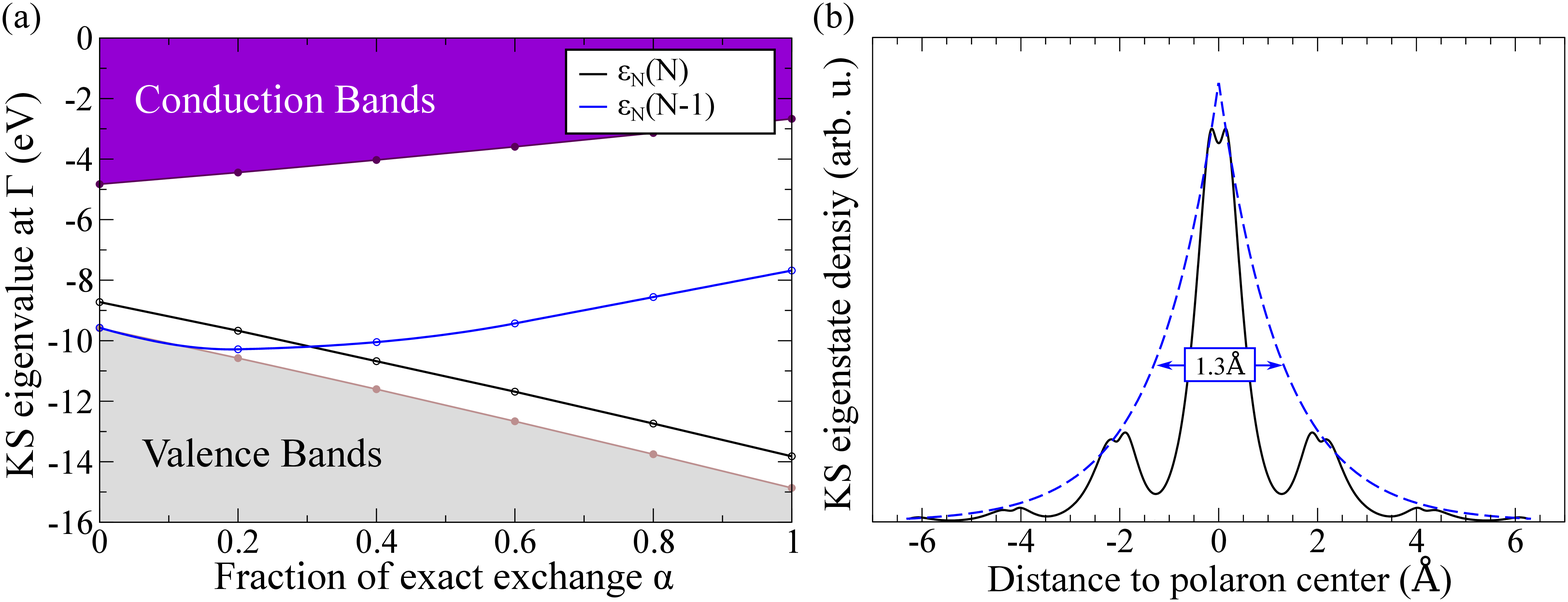}
\caption{(a) The dependence of the hole-polaron level calculated in the charged (blue line) and in the neutral (black line) $3\times3\times3$ supercell of MgO. N corresponds to the number of electrons in the neutral system. Clearly, the black line tracks the VBM for the entire range of $\alpha$, and therefore gives a better description of the polaron level than in the explicitly charged system. (b) Fit for $\rho_m(\vb{r})$ (blue dashed line) of the calculated KS eigenstate density $\rho_d(\vb{r})$ (averaged in [001] direction) of the hole polaron in MgO (black solid line). The obtained polaron radius is 1.3\AA. For details see \ref{app:pekar_polaron}}
\label{fig:ho_lu_dep}
\end{figure}

As a consequence, even with PBE we find a stable self-trapped hole polaron in MgO, which is not the case when charged supercells are used. Also, we find that the polaron level with respect to the band edge ($E_0$), calculated using a neutral supercell, is insensitive to the functional, as can be seen in Fig.~\ref{fig:ho_lu_dep}. A stronger functional dependence of $E_0$ is expected when the character of the polaronic state or states of the band edges are sensitive to the functional.

Using $E_\tx{bind}^0$ for calculating polaron binidng energies has been first implicitly introduced by Zawadski {\em et al.}~\cite{Zawadzki2011}. The independence of $E_\tx{bind}^0$ on the functional has been discussed by Sadigh {\em et al.}~\cite{Sadigh2015}. In their work, Sadigh {\em et al.} have also suggested a way to obtain forces for a polaronic distortion directly using $E_\tx{bind}^0$ potential-energy surface (PES). This facilitates the calculation of accurate elastic response to the excess charge at the level of a hybrid functional, but at the cost of a PBE calculation. 

Thus, using the $E_\tx{bind}^0$ PES instead of charged supercells allows us to significantly reduce the functional dependence. Naively, one may expect that the supercell dependence is also reduced, since only neutral supercell calculations are performed. However, this is not the case. As can be seen from Fig.~\ref{fig:polaron_alpha_1}, panel a, the dependence of $E_\tx{bind}^0$ on the supercell size is much stronger than in the case of charged suercells. This dependence is due to the artificial interaction between ionic relaxation fields in different supercells. Indeed, the $E_\tx{bind}^+ - E_\tx{corr}^\tx{el-st}$ and $E_\tx{bind}^0$ supercell dependence are practically identical and correspond to the long-range part of the electron-phonon interaction potential given by Eq.~\eqref{pekar_potential} in the strong el-ph coupling limit. This understanding allows us to introduce an {\em a posteriori} correction $E_{\rm corr}$ removing the dependence of $E_\tx{bind}^0$ on the supercell size. To remove the artificial interaction terms, we use the approach of Freysoldt {\em et al}.~\cite{Freysoldt2009,Freysoldt2011}, but for a different long-range potential, namely the one given by Eq.~\eqref{pekar_potential}. This new correction scheme relies on the assumption of a strong el-ph coupling, but, as demonstrated below, works reasonably well also for intermediate coupling regimes.

The polaron level $E_0$ also depends on the supercell size. Because of special properties of the small polaron in the adiabatic strong-coupling limit, it is possible to relate the polaron binding energy to the polaron level, in accordance with Pekar's 1:2:3:4 theorem\cite{Lemmens1973} . It follows from the theorem that (see details in App.~\ref{app:pekar_ratio}):
\begin{equation}
E_0 (\infty)  = E_0 (\Omega) + 2 \cdot E_\tx{corr}  \, ,
\label{E0_corr}
\end{equation}
Thus, the correction to $E_0(\Omega)$ in a finite supercell is expected to be about twice as large as for the polaron binding energy calculated using neutral supercells. Indeed, this is what we observe for MgO (see Fig.~\ref{figure_2}, panel a), where the absolute value of the $E_0(L)$ slope is almost exactly twice of the absolute value of the $E_\tx{bind}^0$ slope. For TiO$_2$, the relation between the $E_0(L)$ and $E_\tx{bind}^0$ dependencies deviates from the one derived from Pekar's model (see Fig.~\ref{figure_2}, panel b) due to a weaker electron-phonon coupling, as discussed in detail in Section \ref{sectio:numerical_results}.

In summary, we find that in this approach the dependence on the exchange-correlation approaximation is drastically reduced, but the finite-size effects are significantly more pronounced. However, these effects, caused by the electron-phonon long-range potential (eq.~\eqref{pekar_potential}), can be corrected using the approach of Freysoldt {\em et al.}, but with the potential $V^\tx{lr}_\tx{el-ph}$. This makes possible using moderately sized supercells and semi-local functionals to predict polaron properties, as demonstrated in the next section.

\section{Polarons in rocksalt $\tx{MgO}$ and rutile $\tx{TiO}_2$}
\label{sectio:numerical_results}

Building on the  findings and understanding obtained in the previous sections, we formulate our approach for a reliable calculation of polaron properties:
\begin{enumerate}
\item We obtain the atomic structure of the polaron  using the PBE functional [corresponding to HSE06($\alpha = 0$)] and the approach of Sadigh {et al.}~\cite{Sadigh2015} for each supercell size.
\item HSE06($\alpha=1$) calculations (as a limiting case) are performed for the fixed geometries obtained with PBE. This allows the estimation of the functional dependence for the systems.
\item The polaron binding energies are calculated using Eq.~\eqref{BEPo_neutral}. The finite-size correction for the binding energy is calculated using Eq.~\eqref{eq:ecorr} with the potential from Eq.~\eqref{pekar_potential}. The correction for the polaron level is calculated as twice the correction for the binding energy.
\end{enumerate}
The different sign of the correction for the hole polaron versus the electron polaron (compare panels a and b in Fig.~\ref{figure_2}) is explained by the fact that the equation for the electron affinity has to be used for the electron instead of the ionization potential for the hole. 

We use the hybrid-functional implementation \cite{Levchenko2015} in the all-electron full-potential electronic-structure package FHI-aims \cite{Blum2009,HavuV09,Xinguo/implem_full_author_list}. The evaluation of forces and total energies are computed with FHI-aims using the default \textit{light} settings, to obtain consistent results for all unit cell sizes. As is shown in the Supplementary Information (SI), using default \textit{tight} settings, which are the recommended settings for well-converged calculations, does not affect the results for the smallest supercell. As a demonstration, we apply our new approach to polarons in MgO and rutile TiO$_2$. For the cubic 8-atom MgO unit cell we use a lattice constant of $a=4.211$~\AA~obtained with HSE06 ($\alpha=0.25$), and a $\Gamma$-centered $8\times8\times8$ k-grid. The number of k-points for each direction is scaled down linearly for larger supercell sizes. For the tetragonal 6-atom TiO$_2$ unit cell we use $a=4.64$~\AA~and $c=2.97$~\AA~obtained with the PBE functional and a $9\times9\times15$ k-grid. Due to one more degree of freedom the positions of the atoms are optimized, too, using the PBE functional (for details see SI).

The results for a hole polaron in MgO and an electron polaron in rutile TiO$_2$ are shown in Fig.~\ref{figure_2}. For every supercell size we allow all atoms to relax to obtain the full elastic contribution within the cell. The corrected $E^0_\tx{bind}$ values for each supercell are shown for PBE. Clearly, the supercell-size dependence of $E^0_\tx{bind}$ for both MgO and TiO$_2$ agrees very well with the behavior corresponding to the electron-phonon long-range contribution described by Eq.~\eqref{pekar_potential}. As mentioned above, the Fr{\" o}lich coupling constant $\alpha_\tx{Fr\"ohlich}$ is equal to 4.4 for MgO and 2.2 for TiO$_2$. Thus, MgO is better described by Pekar's model Eq.~\eqref{pekar_potential}, and the size-corrected binding energy practically coincides with the binding energy obtained from a linear extrapolation to the dilute limit. For TiO$_2$, the corrected energy deviates (surprisingly only slightly) from the extrapolated one (within 0.05~eV), reflecting approximations in Pekar's model. Also, the functional dependence of the energies is stronger for TiO$_2$, indicating a larger contribution of the short-range effects to the binding energy. Additionally, we observe that the atomic structure is sensitive to the functional as well demonstrating limitations of obtaining polaron atomic geometries with only the PBE functional, even on the PES corresponding to $E_\tx{bind}^0$, which is much less sensitive to the approximations in the functional than the PES of a charged supercell. This sensitivity is connected to delocalization errors and missing static correlation originated in the d-orbitals. However, the changes in the geometry as a function of $\alpha$ are still small, and we use the configurations of the perfect system obtained with PBE. We find final polaron binding energies in the dilute limit -0.38\dots-0.58~eV for MgO and -0.14\dots-0.41~eV for TiO$_2$, where the range indicates changes in $\alpha$ from 0 to 1. For the polaron level with respect to the band edge we find 1.42\dots1.74~eV for MgO and -0.86\dots-1.44~eV for TiO$_2$. These results remain both qualitatively and quantitatively consistent across a broad range of functionals generated by varying the fraction of exact exchange. This consistency is remarkable when compared to previous theoretical studies, especially for TiO$_2$, since it was either shown that the small polaron formation is expected only for a certain range of a parameter, e.g. for DFT+$U$~\cite{Setvin2014,Deskins2007} or HSE($\alpha$)~\cite{Spreafico2014}, or it was demonstrated only for a specific value of a parameter, e.g. for HSE($\alpha$=0.25)~\cite{Janotti2013}.

To make a connection to experimentally accessible quantities, in particular photoluminescence (PL) measurements, accurately predicting the position of the polaron level is important. Since the quantities obtained with the neutral PES $E^0_\tx{bind}$ are weakly dependent on the underlying functional, the fraction of exact exchange $\alpha$ can be used to tune the gap $E_\tx{gap}$ to recover the experimental band gap. The main PL peak due to the small polaron formation can be expected at:
\begin{equation}
PL = E_\tx{gap}- E_0 \, .
\end{equation}
For MgO the experimental band gap was measured to 7.8-7.9 eV\cite{Madelung2012}, which can be simulated by a fraction $\alpha=0.4$. Based on our HSE06($\alpha = 0.4$) calculations, the PL peak should be at $6.3 \pm 0.1$eV. Unfortunately we could not find any experimental reference for this region of PL. For TiO$_2$ a fraction $\alpha=0.2$ is needed in order to reproduce the experimental band gap of 3.0~eV\cite{Pascual1978}, and the corresponding photoluminescence peak is predicted to be at $2.1\pm 0.1$~eV. This is in good agreement with experimental findings of $PL=2.34$~eV for rutile powders~\cite{Abazovic2006} or direct measurements of the polaron level $E_0=0.7\pm 0.1$~eV with scanning tunneling spectroscopy~\cite{Setvin2014}. We note that the results provided here only represent an upper limit for the polaron level or lower limit for the PL peak, since neither finite-temperature nor non-adiabatic effects are taken into account.

\section{Conclusions}

In this work, we developed a new approach for first-principles modelling of small polarons in materials using DFT supercell calculations. Because on the one hand, the standard charged supercell approach allows us to obtain polaron properties in the dilute limit (for moderately large finite supercells and values of $\epsilon_0$ finite-size errors can be even neglected), but the results strongly depend on the underlying exchange-correlation functional. On the other hand, the approach of Sadigh \textit{et al.}~\cite{Sadigh2015} significantly reduces the dependence on the functional, but, as we demonstrate, introduces a strong dependence on the supercell size. We show that the large finite-size errors in the latter approach are due to constraints imposed on the elastic response to the excess charge by the periodic boundary conditions, and suggest a way to correct the errors for finite supercells. The correction relies on the validity of Pekar's model~\cite{Pekar1946} for the long-range response, based on approximations corresponding to the adiabatic strong (in Fr\"olich's sense) electron-phonon coupling limit. As a result, our approach allows us to obtain polaron properties in the dilute limit and at the same time reduce the exchange-correlation errors, so that even semi-local functionals can be used to reliably estimate polaron level, binding energy, and atomic structure. For more accurate modelling of polaron effects on photoluminescence in materials, the use of hybrid-density functionals or methods beyond DFT, such as the $GW$ approach \cite{Hedin:1965,Rinke/etal:2005,Rinke/etal:2012}, is still necessary.

We apply the developed approach to small polarons in MgO and rutile TiO$_2$. We find that the hole polaron in MgO indeed behaves as Pekar's polaron at the long range, as expected based on the large value of Fr\"olich's constant. For electron polarons in TiO$_2$, our approach also works surprisingly well, considering the weaker electron-phonon coupling in this material. Our all-electron full-potential results support the existence of a small electron polaron in rutile TiO$_2$ in agreement with previous work\cite{Spreafico2014,Setvin2014}.

\section*{Acknowledgments}
This work was supported by the DFG Excellence Cluster ``UniCat'' and the Leibniz ScienceCampus ``GraFOx''.

\appendix

\section{Pekar's Polaron and its relation to KS eigenstates}
\label{app:pekar_polaron}

The objective of the appendix is to show how analytical polaron models are connected to the actual many-body problem treated with DFT. Especially, the relation of the polaron wave function to the highest occupied (ho) or lowest unoccupied (lu) KS state is discussed below.  

Pekar's polaron model~\cite{Pekar1946} evolves from the Fr\"ohlich Hamiltonian \cite{Froehlich1954} in the strong coupling limit, as was shown for example by Devreese\cite{Devreese2016} (cf. also citation in it for original works). In this limit, assuming adiabatic separation of ionic and electronic degrees of freedom, the electron-phonon interaction has the form:
\begin{equation}
V_\tx{el-ph}(\vb{r}) = -\frac{1}{\kappa}\int \frac{|\Phi({\bf r}')|^2}{|{\bf r}-{\bf r}'|} d^3r'\, .
\label{pekar_energy}
\end{equation}
which is the classical response of a polar dielectric to an extended charge distribution. The inverse dielectric constant $\kappa^{-1}=\epsilon_\infty^{-1}-\epsilon_0^{-1}$ describes the polarization of the rigid ions in the medium by the electron or hole. For simplicity, here we assume an isotropic medium (the dielectric response is described by a single constant). Let us regard Eq.~\ref{pekar_energy} as a perturbation of the perfect system $H_\tx{perf}$ -- i.e. the single-electron Hamiltonian, where the electron has been placed at the bottom of the conduction band minimum (CBm) $\phi_\tx{CBm}$ with energy $\varepsilon_\tx{CBm}$ of the non-interacting system (this is the scenario for the electron polaron):
\begin{equation}
H_\tx{perf}\phi_\tx{CBm}=\varepsilon_\tx{CBm}\phi_\tx{CBm}
\end{equation} 
Following the Kohn-Luttinger perturbation theory\cite{Luttinger1955} the solution of:
\begin{equation}
(H_\tx{perf} + V_\tx{el-ph})\Psi =E \Psi 
\end{equation}
in first order is given by:
\begin{eqnarray}
E=\varepsilon_\tx{CBm}+E_0 \nonumber \\
\Psi = \phi_\tx{CBm} \Phi
\label{kohn_luttinger}
\end{eqnarray}
where $E_0$ and $\Phi$ are obtained from the solution of the  effective Hamiltonian of the charge with an effective mass $m^*$~\cite{Luttinger1955} (without taking into account polaronic effects):
\begin{eqnarray}
  (H_\tx{kin,eff}+V_\tx{el-ph})\Phi &&= E_0\Phi \nonumber \\ 
  \left(-\frac{\nabla}{2m^*}-\frac{1}{\kappa}\int \frac{|\Phi({\bf r}')|^2}{|{\bf r}-{\bf r}'|}d^3r'\right)\Phi({\bf r}) &&=  E_0\Phi({\bf r})\, ,
  \label{eq:pekar2}
\end{eqnarray}
with the effective mass $m^*$ from the CBm. With Eq.~\eqref{eq:pekar2} we recover the original problem of Pekar's polaron and $E_0$ is the energy of the bound (polaron) state relative to the conduction-band edge for the case of an electron polaron. 

Eq.~\eqref{eq:pekar2} does not contain microscopic details. However, it can be regarded as describing asymptotic el-ph interaction far away from the localized part of the excess electron charge distribution and, thus, $\Phi({\bf r})$ is the asymptotic solution of Eq.~\ref{eq:pekar2}. According to Eq.~\eqref{kohn_luttinger}, $\Phi({\bf r})$ represents the envelop of the original electronic state $\phi_\tx{CBm}$ and is expected to decay exponentially with distance. The electron KS eigenstate $\phi_\tx{lu}^\tx{DFT}$ corresponding to $\varepsilon_\tx{lu}(N)$ in the DFT calculation at the {\em distorted} (polaron) geometry is the polaron wave function $\Psi$. Thus, the envelop of $\phi_\tx{lu}^\tx{DFT}$ shows the localization of $\rho_d (\vb{r})$ needed for the correction scheme Eq.~\eqref{eq:ecorr} in order to fit $\rho_m (\vb{r})$. An example of $\rho_d (\vb{r})$ calculated with DFT is shown in Fig.~\ref{fig:ho_lu_dep}(b).

%Using the exponential ansatz $\Phi({\bf r})=(\pi r_p^3)^{-1/2}e^{-r/r_p}$, one obtains the solution of Eq.~\eqref{eq:pekar2} by minimizing the virial $J=E_\tx{kin}+1/2E_\tx{el-ph}$ as function of the polaron radius $r_p$\cite{Alexandrov1995}:
%\begin{equation}
%E_0=E_\tx{kin}+E_\tx{el-ph}=-0.146\,E_\tx{h}\,\frac{m^*}{\kappa^2}\, ,
%\end{equation}
%where $E_\tx{h}$ is the Hartree Energy. 

\section{Pekar's 1:2:3:4 theorem}
\label{app:pekar_ratio}

For arbitrary coupling constants in the Fr\"ohlich Hamiltonian:
\begin{equation}
H_\tx{Fr\"ohlich} = H_\tx{kin,eff} + H_\tx{ph} + V_\tx{el-ph},
\end{equation}
with the Hamiltonian of the phonons $H_\tx{ph}$, it has been shown \cite{Lemmens1973} that there exist fixed ratios of the effective kinetic energy $E_\tx{kin,eff}$, lattice distortion (phonon field) energy $\Delta E^\tx{polaron}$, the polaron state energy $E_0$, and the electron-phonon interaction energy $E_\tx{el-ph}$:
\begin{equation}
E_\tx{kin,eff}:\Delta E^\tx{polaron}:-E_0:-E_\tx{el-ph}=1:\eta(\alpha_F):3:4 \, ,
\label{pekar_ratio}
\end{equation}
where $\eta$ depends on the value of Fr{\" o}lich coupling constant $\alpha_F$. In the limit of strong electron-phonon coupling ($\alpha_F \rightarrow \infty$), the polaron energy is dictated by the polarization of the lattice, and $\eta$ approaches 2. From this it follows:
\begin{eqnarray}
E_\tx{bind} &&=  E_\tx{kin,eff}+\Delta E^\tx{polaron}+E_\tx{el-ph}
=  E_\tx{kin,eff}+\frac{1}{2}E_\tx{el-ph} \label{bind} \\
E_0 &&= E_\tx{kin,eff} + E_\tx{el-ph} \label{e0}
\end{eqnarray}
Eqs.~\eqref{bind} and~\eqref{e0} clearly show the dependence of the binding energy and the polaron level on the energy of the el-ph interaction. The latter energy is the one that remains to be corrected for the artificial supercell interactions, and thus the correction for the polaron level has to be twice of the correction for the binding energy, which leads to Eq.~\eqref{E0_corr}.

However, these ratios Eq.~\eqref{pekar_ratio} are only based on an effective single-particle model (Eq.~\ref{eq:pekar2}). In our microscopic (DFT) model, additional (short-range) contributions to the energy components and deviations of $\eta\leq2$ lead to violation of the above ratios. In particular for TiO$_2$ the ratios are not preserved. However, for MgO, where the Fr{\" o}lich constant is 4.4, indicating indeed a strong electron-phonon coupling, the ratios are close to the ones found by Pekar, and the polaron level and binding energies calculated from the model are close to the ones from DFT calculations, as described in the text.

\section{Freysoldt \textit{et al.} correction scheme for finite-size effects in a nutshell}\label{app:Freysoldt}
A repeating point in this paper is the correction of finite-size effects for supercell calculations. For completeness we present the main ideas of the correction scheme proposed by Freysoldt \textit{et al.}\cite{Freysoldt2009,Freysoldt2011}. Starting point is the simulation of a charged point defect in an otherwise pristine crystal causing a localized excess-charge distribution $\rho_d$. It is assumed that for a sufficient large supercell the quantum nature of the defect is simulated properly and only long-ranged interactions do affect the defect potential in neighboring cells. If the long-range potential possess a Fourier-transformation, e.g. as shown here for $V^\tx{lr}_\tx{el-st}$ (eq.~\eqref{elst_potential}) and $V^\tx{lr}_\tx{el-ph}$ (eq.~\eqref{pekar_potential}), then, it is possible to correct the biased energies \textit{a posteriori}. For this, to have simple evaluable sums and integrals Freysoldt \textit{et al} suggest to model $\rho_d$ with a simple isotropic function $\rho_m$, such as an exponential or a Gaussian (the fitting of $\rho_d$ by $\rho_m$ is demonstrated in App.\ref{app:pekar_polaron}). The actual detailed excess-charge distribution is not necessary to know and would change the correction only negligibly. (As Freysoldt \textit{et al.} in their original paper note it is not even important to imitate the proper localization of $\rho_d$ as long as the distribution is well-localized within the supercell.) With this, it is possible to evaluate the lattice sum of the long-range potential (i.e. the potential energy due to their periodic arrangement):
\begin{equation}
E_\tx{latt} =\frac{1}{\Omega}\sum_{{\bf G}\neq {\bf 0}} V^\tx{lr}(\vb{G}) q_m(\vb{G})
\label{eq:latt_sum}
\end{equation}
(for the detailed nomenclature see main text), where $V(\vb{G})$ is the Fourier-transform of the long-range potential, and the sum runs over all reciprocal lattice vectors $|\vb{G}|<G_\tx{cut}$. The cut-off $G_\tx{cut}$ has to be chosen carefully to ensure convergence of the sum. Eq.~\ref{eq:latt_sum} is the artificial energy, which has to be removed from the regarded energy (e.g. the polaron binding energy or level). What is missing is the long-range energy of the isolated defect. This is easily calculated by:
\begin{equation}
E_\tx{iso}=\frac{1}{(2\pi)^3} \int V(\vb{k}) q_d(\vb{k}) \tx{d}\vb{k}\,
\end{equation}
and the total correction is given by $E_\tx{corr}=E_\tx{latt}-E_\tx{iso}$. To obtain the desired energy in its dilute $E_\infty$ limit the correction $E_\tx{corr}(\Omega)$ has to be removed from the energy $E(\Omega)$ calculated in the supercell of size $\Omega$:
\begin{equation}
E_\infty = E(\Omega) - E_\tx{corr}(\Omega) + q\Delta V.
\end{equation}
The last term $q\Delta V$ is the so-called alignment term and has to be considered for the following reasons: First, usually $E(\Omega)$ is calculated with respect to a reference system, often the pristine bulk system. Due to defect or the charge there might be difference in the potentials for the defect system and the pristine system even far away from the defect center. This difference can be obtained by aligning the electrostatic potentials (or Hartree potentials). Second, the absolute position of the long-range potential calculated from $\rho_m$ might not be equal to the one from the original $\rho_d$. This difference must be aligned, too. Hence, in general the term $q\Delta V$ should include these two contributions.  

\bibliographystyle{unsrt}

\end{document}